\documentclass[article]{llncs}
\usepackage{graphicx}
\usepackage{amsmath,amsthm,amssymb}

\usepackage{xcolor}
\usepackage{colortbl}

\usepackage{wrapfig}

\usepackage{multirow}
\usepackage{verbatim} 
\usepackage{booktabs}

\usepackage{algorithm, algorithmic}

\usepackage[breaklinks=true,hidelinks]{hyperref} %<--- Load after everything else

\begin{document}
\title{Enhanced OpenMP Algorithm to Compute All-Pairs Shortest Path on x86 Architectures}
\author{Sergio Calderón\inst{1,2}
\and Enzo Rucci\inst{1,3}\orcidID{0000-0001-6736-7358}\thanks{Corresponding author.} \and
Franco Chichizola\inst{1}\orcidID{0000-0001-8857-6343} 
%\and Marcelo Naiouf\inst{1}\oorcidID{0000-0001-9127-3212}
}
%
%\authorrunning{E. Rucci et al.}
\institute{III-LIDI, Facultad de Informática, UNLP – CIC. \\La Plata (1900), Bs As, Argentina \\
\email{\{scalderon,erucci,francoch\}@lidi.info.unlp.edu.ar} \\
\and
Becario de Entrenamiento, CIC
\and
Comisión de Investigaciones Científicas (CIC). \\La Plata (1900), Bs As, Argentina
 }
\maketitle              % typeset the header of the contribution
\begin{abstract}

Graphs have become a key tool when modeling and solving problems in different areas. The Floyd-Warshall (FW) algorithm computes the shortest path between all pairs of vertices in a graph and is employed in areas like communication networking, traffic routing, bioinformatics, among others. However, FW is computationally and spatially expensive since it requires $O(n^3)$ operations and $O(n^2)$ memory space. As the graph gets larger, parallel computing becomes necessary to provide a solution in an acceptable time range. In this paper, we studied a FW code developed for Xeon Phi KNL processors and adapted it to run on any Intel x86 processors, losing the specificity of the former. To do so, we verified one by one the optimizations proposed by the original code, making adjustments to the base code where necessary, and analyzing its performance on two Intel servers under different test scenarios. In addition, a new optimization was proposed to increase the concurrency degree of the parallel algorithm, which was implemented using two different synchronization mechanisms. The experimental results show that all optimizations were beneficial on the two x86 platforms selected. Last, the new optimization proposal improved performance by up to 23\%.

%En la actualidad, uno de los principales desafíos de los sistemas de cómputo de alto rendimiento consiste en mejorar su rendimiento manteniendo el consumo energético en niveles aceptables. En ese sentido, una estrategia consolidada consiste en emplear aceleradores como las GPUs o los procesadores many-core Intel Xeon Phi. En este trabajo, se describen y comparan dispositivos de las arquitecturas NVIDIA Pascal e Intel Xeon Phi Knights Landing. Seleccionando el algoritmo de Floyd-Warshall como caso representativo de aplicaciones de grafos y limitadas por memoria, se desarrollaron implementaciones optimizadas con el fin de analizar y comparar el rendimiento y la eficiencia energética en ambos dispositivos. Contrariamente a lo considerado en el análisis preliminar, se encontró que el rendimiento de ambos dispositivos resultó comparable mientras que, respecto a la eficiencia energética, el Xeon Phi se mostró superior.

\keywords{ Floyd-Warshall \and Multicore \and APSP \and Xeon \and Xeon Phi Knights Landing \and Core \and OpenMP }
\end{abstract}

\begin{center}
\texttt{This version of the contribution has been accepted for publication, after peer review (when applicable) but is not the Version of Record and does not reflect post-acceptance improvements, or any corrections. The
Version of Record is available online at: \url{https://doi.org/10.1007/978-3-031-62245-8\_4}. Use of this
Accepted Version is subject to the publisher’s Accepted Manuscript terms of use
\url{https://www.springernature.com/gp/open-research/policies/accepted-manuscript-terms}}
\end{center}

\clearpage

\section{Introduction}

The Floyd-Warshall (FW)~\cite{Floyd,Warshall} algorithm computes the shortest path between all pairs of vertices in a graph and is employed in areas like communication networking~\cite{Floyd_networking}, traffic routing~\cite{floyd_traffic}, bioinformatics~\cite{floyd_bioinformatics}, among others. However, FW is  computationally and spatially expensive since it requires $O(n^3)$ operations and $O(n^2)$ memory space, where $n$ is the number of vertices in a graph. As the graph gets larger, parallel computing becomes necessary to provide a solution in an acceptable time frame. This is why the scientific community has made multiple efforts for this purpose~\cite{floyd_blocked1,floyd_blocked2,floyd_han_kang,floyd_han,floyd_mc,floyd_mpi,floyd_hi,floyd_spark}. Focusing on Intel Xeon Phi platforms, Rucci \textit{et al.}~\cite{RucciCACIC2017} explored its use to accelerate FW in the first generation (KNC, Knights Corner), while Costi \textit{et al.}~\cite{costi2020aceleracion} extended it to the second (Knights Landing, KNL). 

In \cite{CalderonCACIC2023}, we studied the code developed by ~\cite{costi2020aceleracion} and adapted it to run on Intel x86 processors, losing the specificity of the Xeon Phi KNL. To do so, we verified one by one the optimizations proposed by~\cite{costi2020aceleracion}, making adjustments to the base code where necessary, and analyzing its performance on two Intel servers under different test scenarios. This paper is an extended and thoroughly revised version of~\cite{CalderonCACIC2023}. The work has been extended by providing: 
\begin{itemize}
    \item The proposal of a new optimization for the parallel FW algorithm (\textit{intra-round parallelism}) and its implementation using different synchronization mechanisms (semaphores vs. condition variables).
    \item A performance analysis of the proposed optimization on 2 different Intel x86 servers.
    \item The creation of a public git repository with the different
codes developed for this paper~\footnote{\url{https://bit.ly/cacic23-fw}}.
\end{itemize}

The rest of the paper is organized as follows. Section~\ref{sec:back} introduces the background of this work.
Section~\ref{sec:trabajo} details the adaptation process and the new optimizacion proposal. In Section ~\ref{sec:resultados}, performance
results are presented and finally, in Section~\ref{sec:conclusiones}, conclusions and some ideas for
future research are summarized.

\section{Background}
\label{sec:back}

\subsection{Intel Xeon Phi}

Xeon Phi is the brand name that Intel used for a series of many-core, HPC-oriented processors. In 2012, Intel launched the first Phi generation (KNC) with 61 x86 pentium cores (4 hardware threads per core) equipped with a 512-bit vector unit (VPU) each. In contrast to KNC co-processors connected via a PCI Express bus to the host, the second Phi generation (KNL) can act as self-boot processors. KNL processors feature a large number of cores with hyper-threading support, the incorporation of AVX-512's 512-bit vector instructions, and the integration of a high-bandwidth memory (HBM), among others~\cite{KNLbook}. The latest generation (Knight Mills, KNM) was released in late 2017, being a KNL variant with specific instructions for deep machine learning. Finally, Intel announced that it would discontinue the Xeon Phi series in 2018 to focus on the development of graphics boards (GPUs)~\footnote{\url{https://hardzone.es/2018/07/25/intel-adios-xeon-phi-reemplazados-tarjetas-graficas/}}.

\subsection{Intel Xeon and Core}

Currently, Intel offers two processor segments in the x86 family: Xeon and Core. Intel Xeon processors are designed for enterprise and server tasks that require high-performance computing and reliability, while Intel Core processors are ideal for general-purpose use, including gaming, office applications and multimedia entertainment. 

In terms of architectural features, Xeon processors tend to have more cores, specific technologies for virtualization and security, support for multi-socket configuration, among other advanced features. In opposite sense, Core processors tend to have higher frequencies, lower power consumption and lower price.

The choice between Xeon and Core will depend on the specific application needs, budget and user preferences.

\subsection{FW Algorithm}

\begin{wrapfigure}{r}{0.5\textwidth}
 \begin{minipage}{0.5\textwidth}
\begin{algorithmic}
\FOR{$k \gets 0$ to $N-1$}
    \FOR{$i \gets 0$ to $N-1$}
        \FOR{$j \gets 0$ to $N-1$}
            \IF {$D_{i,j}$ $\geq$ $D_{i,k}$ + $D_{k, j}$} 
                \STATE $D_{i,j}$ $\gets$ $D_{i,k}$ + $D_{k, j}$
                \STATE $P_{i,j}$ $\gets$ $k$
            \ENDIF
        \ENDFOR
    \ENDFOR
\ENDFOR
\end{algorithmic}

  \end{minipage}
\caption{Pseudocode of the FW algorithm}
\label{fig:algfw}

\end{wrapfigure}

The pseudocode of FW is shown in Fig.~\ref{fig:algfw}. Given a graph \textit{G} of \textit{N} vertexes, FW receives as input a dense \textit{N}$\times$\textit{N} matrix $D$ that contains the distances between all pairs of vertexes from \textit{G}, where $D_{i,j}$ represents the distance from node \textit{i} to node \textit{j}~\footnote{If there is no path between nodes \textit{i} and \textit{j}, their distance is considered to be infinite (usually represented as the largest positive value)}. FW computes \textit{N} iterations, evaluating in the \textit{k}-th iteration all possible paths between vertexes \textit{i} and \textit{j} that have \textit{k} as the intermediate vertex. As a result, it produces an updated matrix \textit{D}, where $D_{i,j}$ now contains the shortest distance between nodes \textit{i} and \textit{j} up to that step. Also, FW builds an additional matrix \textit{P} that records the paths associated with the shortest distances.

%\begin{wrapfigure}{r}{0.5\textwidth}
%    \includegraphics[width=.95\linewidth]{figs/fw.png}
%    \caption{Pseudocódigo del algoritmo de FW}
%    \label{fig:fw}
%\end{wrapfigure}

\subsubsection{Blocked FW Algorithm.}

At first glance, the nested triple loop structure of this algorithm is similar to that of dense matrix multiplication (MM). However, since read and write operations are performed on the same matrix, the three loops cannot be freely exchanged, as is the case with MM. Despite this, the FW algorithm can be computed by blocks under certain conditions~\cite{floyd_blocked2}.

The blocked FW algorithm (BFW) divides matrix \textit{D} into blocks of size $BS\times BS$, totaling $(N/BS)^2$ blocks. Computation is organized in $R=N/BS$ rounds, where each round consists of 4 phases ordered according to the data dependencies between the blocks:

\begin{enumerate}
    \item Phase 1: Update the $D^{k,k}$ block because it only depends on itself.
    \item Phase 2: Update the blocks in row \textit{k} of blocks ($D^{k,*}$) because each of these depends on itself and on $D^{k,k}$. 
    \item Phase 3: Update the blocks in column \textit{k} of blocks ($D^{*,k}$) because each of these depends on itself and on $D^{k,k}$.  
    \item Phase 4: Update the remaining $D^{i,j}$ blocks of the matrix because each of these depends on blocks $D^{i,k}$ and $D^{k,j}$  on its row and column of blocks, respectively.
\end{enumerate}

\begin{figure}[t]
    \centering
    \includegraphics[width=1\textwidth]{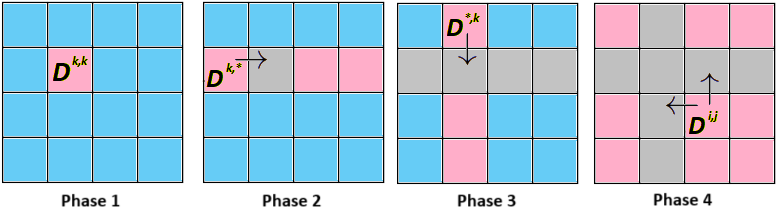}
    \caption{BFW computation phases and block dependencies}
    \label{fig:phases}
\end{figure}

Figure~\ref{fig:phases} shows each of the computation phases and the dependencies between blocks. The pink squares represent blocks that are being computed, gray squares are those that have already been processed, and sky-blue squares are the ones that have not been computed yet. Last, arrows show the dependencies between blocks for each phase.

\subsection{Base code}

As base code, we used the one from~\cite{costi2020aceleracion}, which is a FWB algorithm specifically developed for Intel's Xeon Phi KNL processors. The following is a description of the different optimizations considered in the previous work:

\begin{itemize}

    \item \textbf{Opt-0: Multi-threading}. A multi-threaded version is obtained using OpenMP. In Phases 2 to 4, the blocks are distributed among the different threads utilizing the \texttt{for} directive with \texttt{dynamic} scheduling. In the case of Phase 1, since it consists of a single block, the iterations within it are distributed among the threads.

    \item \textbf{Opt-1: MCDRAM}. Since this is a bandwidth-limited application, using this special memory is greatly beneficial. Executions are done using the \texttt{numactl -p} command to use the DDR memory as an auxiliary one.

    \item \textbf{Opt-2: SSE vectorization}. Using the OpenMP \texttt{simd} directive, the operations of the innermost loop are vectorized when computing each block. Typically, compilers use the 128-bit SSE instruction set by default, which allows the CPU to pack 2 double-precision multiply-add operations (4 flops) or 4 single-precision multiply-add operations (8 flops). 

    \item \textbf{Opt-3: AVX2 vectorization}. AVX2 doubles the number of simultaneous operations concerning SSE. Thus, the CPU is guided to use 256-bit AVX instructions by adding the \textit{-xAVX2} flag to the compilation process (if supported).

    \item \textbf{Opt-4: AVX-512 vectorization}. As the previous case, the \textit{-xMIC-AVX512} flag is included to use 512-bit AVX512 extensions. In this way, the CPU can compute 8 double-precision multiply-add operations (16 flops) or 32 single-precision multiply-add operations (64 flops). 

    \item \textbf{Opt-5: Data alignment}. \texttt{\_mm\_malloc()} allocates aligned blocks of memory, i.e.,  data is stored aligned to the beginning of each cache line. In this way, subsequent read and write operations are optimized.     

    \item \textbf{Opt-6: Branch prediction}.  The distance comparison is a hotspot of FW.  By including the built-in \texttt{\_\_builtin\_expect} compiler macro, \texttt{if} statement branches can be better predicted. The more the scheduler gets right, the more instruction-level parallelism the processor can exploit.

    \item \textbf{Opt-7: Loop unrolling}. By fully unrolling the innermost loop and loop \textit{i} only once.
    
    \item \textbf{Opt-8: Thread affinity}. Threads are distributed among cores according to the variable \texttt{KMP\_AFFINITY}.
  Different affinity types (\textit{balanced}, \textit{compact} or \textit{scatter}) and granularities (\textit{fine}, \textit{core} or \textit{tile}) can be specified. %In Xeon Phi KNL, the optimal configuration was \textit{fine, balanced}.  

\end{itemize}

\section{Implementation}
\label{sec:trabajo}

\subsection{Code adaptation to x86 architectures}

The code base from Section 2 was adapted for its execution on the two x86 servers (see Table~\ref{tab:platforms}). In both cases, the Intel ICC compiler was used, which is part of the oneAPI suite (v2021.7.1).

\begin{table}[tb]
\centering
\caption{Experimental platforms}
\label{tab:platforms}
\resizebox{\columnwidth}{!}{%
\begin{tabular}{@{}ccc@{}}
\toprule
\textbf{ID}            & \textbf{Core i5}     & \textbf{Xeon Platinum}      \\ \midrule
Processor              & Intel Core i5-10400F & 2$\times$Intel Xeon Platinum 8276L \\
Cores (ht)              & 6 (12)               & 56 (112)                    \\
Clock Frequency (base) & 2.9Ghz               & 2.2Ghz                      \\
RAM memory             & 32 GB                  & 250 GB                         \\
OS                     & Debian 11            & Ubuntu 20.04 LTS            \\ \bottomrule
\end{tabular}%
}
\end{table}

The adjustments carried out to the different code versions are detailed below:

\begin{itemize}
    \item  \textbf{Opt-0:} No changes are required in this version.

    \item \textbf{Opt-1:} None of the platforms has MCDRAM memory, so the command \textit{numactl} is ruled out.

    \item \textbf{Opt-2 / Opt-3:} No changes are required in these version.

    \item \textbf{Opt-4:} The associated flag is replaced by the one recommended for the Xeon Platinum (from \textit{-xMIC-AVX512} to \textit{-xCORE-AVX512}); at the same time,  it is discarded on Intel Core i5 because that extension set is not available on that processor.

    \item \textbf{Opt-5:} The Xeon Platinum keeps SIMD\_WIDTH = 512; while Core i5 reduces to 256 because AVX2 is the widest vectorization set available.

    \item \textbf{Opt-6 / Opt-7:} No changes are required in these versions.

    \item \textbf{Opt-8:} No changes are required in this version. The best affinity and granularity configuration is empirically selected.
    
\end{itemize}

\subsection{Opt-9: Intra-round concurrency}

\begin{figure}[hb]
    \centering
    \includegraphics[width=0.9\columnwidth]{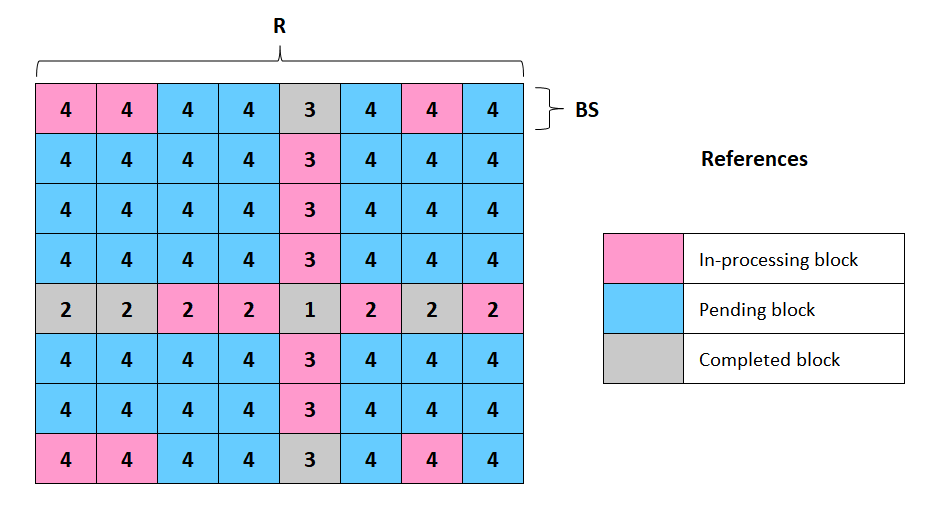}
    \caption{Example of optimization opportunity when $R = 8$ and $k = 4$}
    \label{fig:ph4-temprana}
\end{figure}

This section describes a new optimization proposal, which seeks to increase the concurrency in block computation. In the FWB algorithm, phase 4 of each round must wait for the end of phases 2 and 3 above. However, as phases 2 and 3 progress, some blocks of phase 4 could already be computed (those whose dependencies have already been resolved), without waiting for the end of phases 2 and 3. Fig.~\ref{fig:ph4-temprana} illustrates this improvement opportunity, where the computed blocks are shown in gray (five blocks of phase 2-3), and those in processing are shown in pink (six blocks of phase 4). This possible optimization becomes more relevant when \textit{T} is \textit{large} and \textit{BS} is \textit{small}. 

From a coding perspective, this proposal requires finer-grained synchronization than OpenMP directives can provide; thus, it must work at the Pthreads level. In this sense, two possible implementations for this idea are described below, differing in the synchronization mechanism that it is used.

\subsubsection{Semaphores}

The first version employs semaphores to synchronize threads, through the POSIX library (\texttt{semaphore.h}).
A semaphore matrix of dimension $R \times R$ is added and initialized to zero. Each cell represents a block of the $D$ matrix. The computation of a phase 4 block is conditioned by $d$ \texttt{sem\_wait} operations, where $d$ is the number of dependencies it possesses (in this case $d=2$). The threads that compute the dependent blocks of phases 2 and 3 are responsible for performing the corresponding \texttt{sem\_post} operations.

 After a phase 2 block is computed, a \texttt{sem\_post} is performed on each semaphore of its same column $j$ (except on its own position). Similarly, after processing a phase 3 block, a \texttt{sem\_post} is performed on each semaphore of its same row $i$  (except on its own position). To compute a $D_{i,j}$ block of phase 4, two \texttt{sem\_wait} operations are performed on its own position $(i,j)$. In this way, the dependencies of phase 4 are respected at the block level.
 
\begin{figure}[H]
    \centering
    \includegraphics[width=0.97\columnwidth]{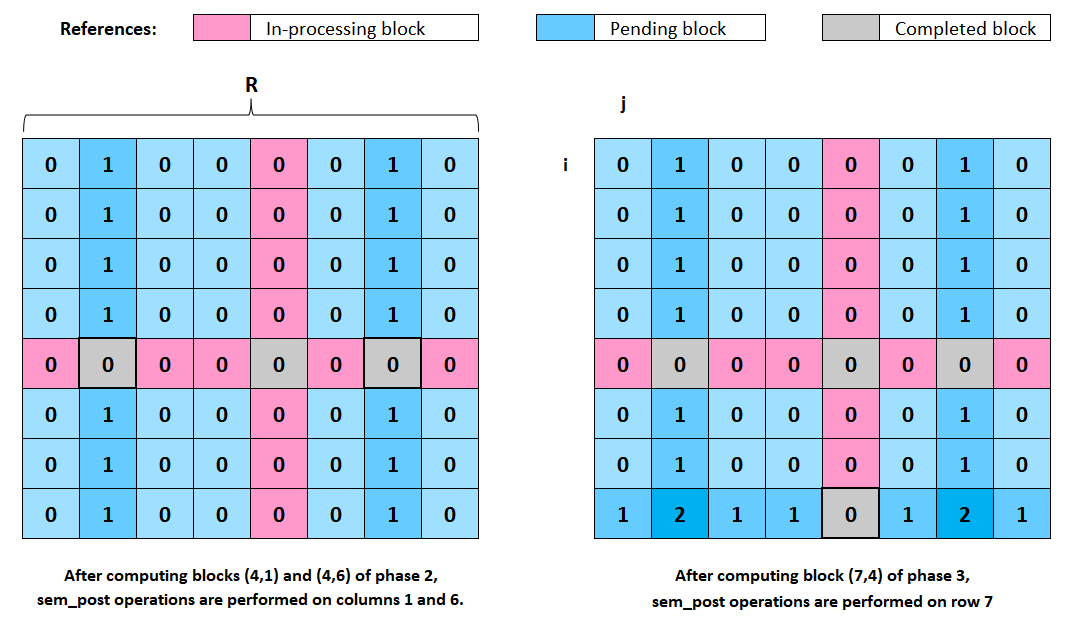}
    \caption{Example of semaphores values during the execution of phases 2 and 3}
    \label{fig:o9-sem}
\end{figure}

When a round ends, all semaphores are set to zero again. The Fig.~\ref{fig:o9-sem} shows the value of the semaphores before \texttt{sem\_wait} operations by phase 4.

\subsubsection{Condition variables}

The second version employs condition variables (cv) to synchronize threads, which are included in Pthreads library.
Three additional data structures are required for this version: a cv matrix ($CV$), a \texttt{mutex} matrix ($M$), and an integer matrix ($F$); all containing $R \times R$ elements.
$M$ is necessary to operate over $CV$ while
each cell $F_{i,j}$ indicates the number of pending dependencies to enable the computation of a block $D_{i,j}$, located in phase 4.

\begin{figure}[H]
    \centering
    \includegraphics[width=0.97\columnwidth]{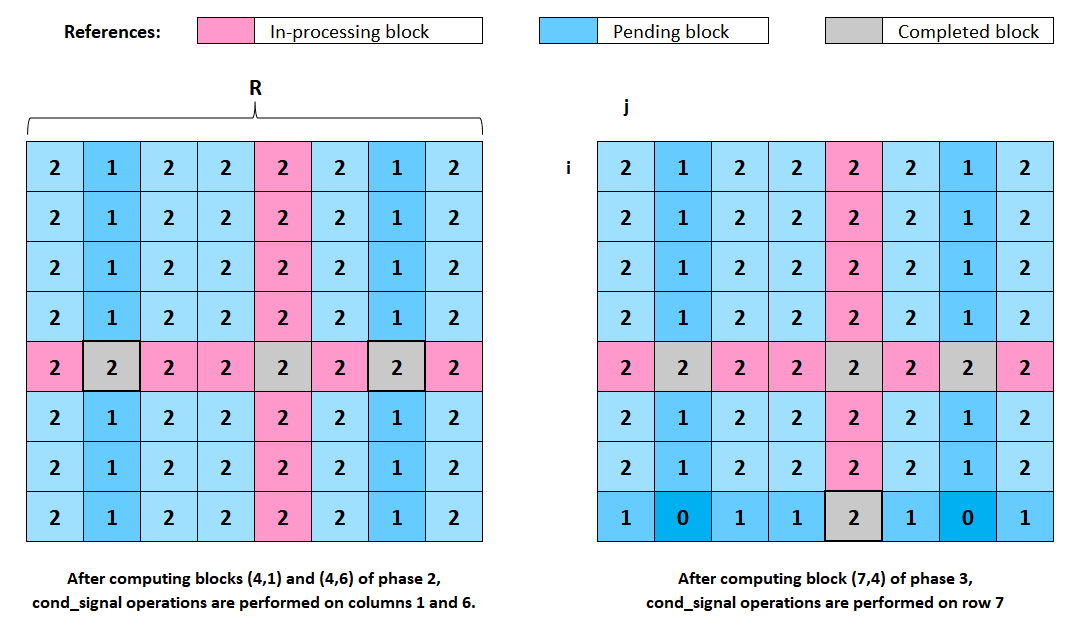}
    \caption{Example of $F$ values during the execution of phases 2 and 3}
    \label{fig:o9-cond}
\end{figure}

As in the previous case, $d=2$ in phase 4. Therefore, $F$ is initialized with this value for all its cells in each round. A thread will only continue when $F_{i,j}=0$; otherwise, it will remain suspended. 
After computing a phase 2 block, the remaining $R-1$ positions of the $F$ matrix in the same $j$ column are first decremented by one (ensuring mutual exclusion). Then, a \texttt{cond\_signal} operation is performed on the associated condition variables.
Analogously, after processing a phase 3 block, one unit is subtracted from the positions of the same column $i$ in $F$ (again, ensuring mutual exclusion). Then, a \texttt{cond\_signal} is performed on the corresponding condition variables. Fig.~\ref{fig:o9-cond} illustrates $F$ for the same case analyzed in Fig.~\ref{fig:o9-sem}.

%% justificación de la matriz F

The matrix $F$  must necessarily be included in this solution. When a \texttt{cond\_signal} is performed on a cv, it will wake up the first thread in the queue (if any); otherwise, it will have no effect. Since the \texttt{cond\_wait} operation always suspends a thread, it should only be called when pending dependencies exist. Additionally, each $F$ update must be placed in a critical section, due to modification conflicts between phase 2 and 3 blocks (different threads could update the same cell).
\section{Experimental Results}
\label{sec:resultados}

\subsection{Experimental Design}

The experiments were carried out on the experimental platforms described in Table~\ref{tab:platforms}. The tests considered the variation of workload ($N$ = \{4096, 8192, 16384\}), data type (\textit{float}, \textit{double}), number of threads ($T_{Core}$ = \{6,12\}, $T_{Xeon}$=\{56,112\}) and block size ($BS$ = \{32, 64, 128, 256\}), where necessary.

All code versions work with the same input data, considering 30\% of null values in the distance matrix.
Each particular test was executed 8 times to minimize fluctuation, and the performance was computed based on the average of these multiple runs. Last, all code versions are available in a public web repository: \url{https://bit.ly/cacic23-fw}.

\subsection{Experimental results of x86 adaptation}

The GFLOPS (billion FLOPS) metric is used to evaluate performance: GFLOPS=$\frac{2 \times N^3}{t \times 10^9}$, where \textit{t} is execution time (in seconds), and the factor 2 represents the number of floating point operations in the inner loop of any FW algorithm.

Considering an intermediate input size $(N = 8192)$ and data type=\textit{float}, the results obtained for each platform are presented in Tables~\ref{tab:i5-tabla} (Core i5) and \ref{tab:xeon-tabla} (Xeon Platinum). It can be seen that the best performance for both servers is achieved using one thread per logical core and $BS=$128 from \textit{Opt-3} onwards.

%\begin{figure}[tb]
%    \centering
%    \includegraphics[scale=0.7]{camera-ready/gflops-i5-8192-float.png}
%    \caption{GFLOPS promedio en Core i5 para N=8192 y tipo \textit{float}}
%    \label{fig:i5-tabla-1f}

%\end{figure}

% \usepackage{booktabs}
% \usepackage{multirow}
% \usepackage{graphicx}
\begin{table}[hb]
\centering
\caption{Performance (average GFLOPS) on Core i5 when N=8192 and datatype=float}
\label{tab:i5-tabla}
%\resizebox{\columnwidth}{!}{%
\begin{tabular}{@{}cc|ccccccc@{}}
\toprule
\textbf{T} & \textbf{BS} & \textbf{Opt-0} & \textbf{Opt-2} & \textbf{Opt-3} & \textbf{Opt-5} & \textbf{Opt-6} & \textbf{Opt-7} & \textbf{Opt-8} \\ \midrule
\multirow{4}{*}{6}  & 32  & 12.35 & 20.62 & 61.92 & 65.95 & 78.89 & 103.11 & 103.44  \\
                     & 64  & 14.37 & 19.38 & 73.58 & 78.41 & 95.47 & 136.61 & 137.42 \\
                     & 128 & 19.99 & 23.40 & 77.06 & 83.50 & 104.07 & 146.66 & 146.70 \\
                     & 256 & 20.34 & 24.61 & 55.09 & 56.99 & 86.83 & 110.86 & 111.34 \\ \midrule
\multirow{4}{*}{12} & 32  & 16.24 & 28.15 & 66.85 &  70.26 & 87.15 & 112.24 & 112.65 \\
                     & 64  & 19.99 & 25.52 & 78.29 & 81.36 & 101.56 & 146.24 & 147.05   \\
                     & 128 & 21.09 & 29.92 & \cellcolor{lightgray}{78.41} &  \cellcolor{lightgray}{82.82} & \cellcolor{lightgray}{107.65} & \cellcolor{lightgray}{154.22} & \cellcolor{lightgray}{154.29} \\
                     & 256 & \cellcolor{lightgray}{21.22} & \cellcolor{lightgray}{32.13} & 60.51 &  62.12 & 94.19 & 110.68 & 110.80 \\
                     \bottomrule
\end{tabular}%
%}
\end{table}

%\begin{figure}[tb]
%    \centering
%    \includegraphics[scale=0.7]{camera-ready/gflops-xeon-8192-float.png}
%    \caption{GFLOPS promedio en Xeon Platinum para N=8192 y tipo \textit{float}}
%    \label{fig:xeon-tabla-1f}
%\end{figure}

\begin{table}[th]
\centering
\caption{Performance (average GFLOPS) on Xeon Platinum when N=8192 and datatype=float}
\label{tab:xeon-tabla}
%\resizebox{\columnwidth}{!}{%
\begin{tabular}{@{}cc|cccccccc@{}}
\toprule
\textbf{T} & \textbf{BS} & \textbf{Opt-0} & \textbf{Opt-2} & \textbf{Opt-3} & \textbf{Opt-4} & \textbf{Opt-5} & \textbf{Opt-6} & \textbf{Opt-7} & \textbf{Opt-8} \\ \midrule
\multirow{4}{*}{56}  & 32  & 62.03 & 111.06 & 116.07 & 185.12 & 222.85 & 346.40 & 374.78 & 494.90  \\
                     & 64  & 62.95 & 81.35 & 111.03 & 189.76 & 230.82 & 472.82 & 480.55 & 710.57 \\
                     & 128 & 87.15 & 110.83 & 132.55 & 215.53 & 334.13 & 507.52 & 559.67 & 831.70 \\
                     & 256 & 87.90 & 111.65 & 112.06 & 185.94 & 232.88 & 441.03 & 444.93 & 641.24 \\ \midrule
\multirow{4}{*}{112} & 32  & 79.23 & 157.39 & 224.92 & 261.30 & 273.38 & 422.60 & 440.06 & 463.17 \\
                     & 64  & 78.10 & 101.26 & 402.67 & 447.31 & 491.73 & 587.05 & 611.58 & 664.20   \\
                     & 128 & 124.09 & 163.40 & \cellcolor{lightgray}{425.86} & \cellcolor{lightgray}{489.67} & \cellcolor{lightgray}{593.25} & \cellcolor{lightgray}{700.64} & \cellcolor{lightgray}{766.82} & \cellcolor{lightgray}{866.31} \\
                     & 256 & \cellcolor{lightgray}{124.47} & \cellcolor{lightgray}{169.32} & 336.85 & 354.80 & 372.23 & 432.98 & 456.85 & 470.91 \\
                     \bottomrule
\end{tabular}%
%}
\end{table}

\begin{figure}[th!]
    \centering
    \includegraphics[width=0.9\columnwidth]{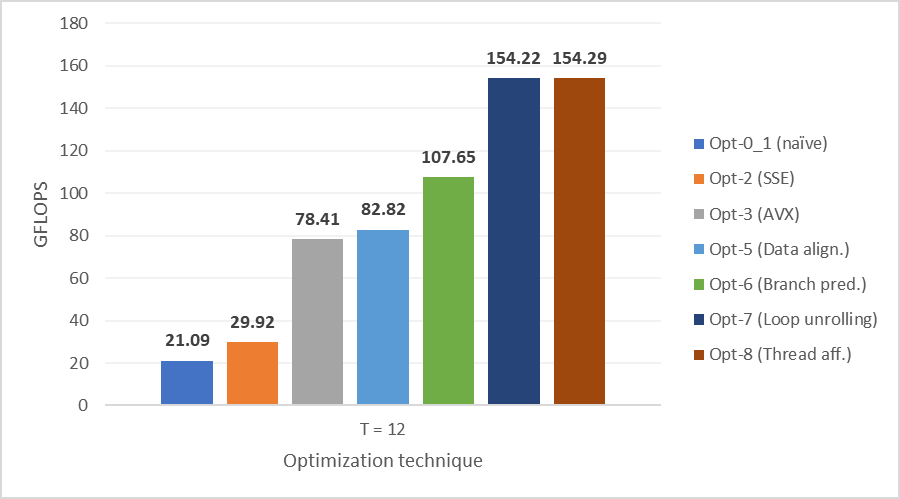}
    \caption{Performance on Core i5 when N=8192 and data type=\textit{float}  (optimal configuration for each version)}
    \label{fig:i5-barras}
\end{figure}

\begin{figure}[th!]
    \centering
    \includegraphics[width=0.9\columnwidth]{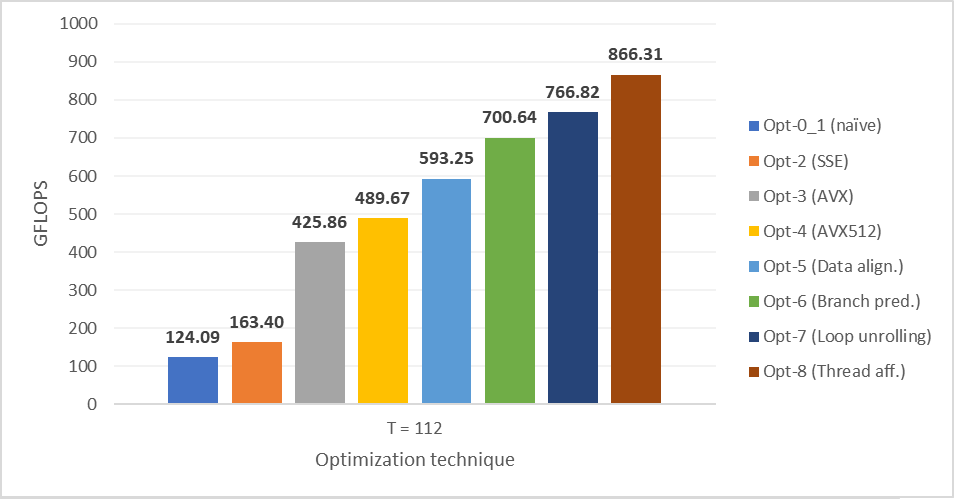}
    \caption{Performance on Xeon Platinum when N=8192 and data type=\textit{float} (optimal configuration for each version)}
    \label{fig:xeon-barras}

\end{figure}

 Table~\ref{tab:mejoras-incrementales} presents the improvement factor for each version over its predecessor, including their comparison with~\cite{costi2020aceleracion}. Fig.~\ref{fig:i5-barras} and~\ref{fig:xeon-barras} show the performance achieved using the aforementioned optimal configuration on the Core i5 and Xeon Platinum machines, respectively. It can be seen that each optimization proposal effectively leads to an increase in the GFLOPS obtained on both machines. The largest improvement is achieved in the \textit{Opt-3} version by vectorizing with \textit{AVX-2} (approximately 2.6$\times$). Then, when comparing the widest vectorization option versus the one that does not vectorize (\textit{Opt-0}), a total improvement of 3.96$\times$ and 3.72$\times$ are obtained on the Xeon Platinum and the Core i5 platforms, respectively. On its behalf, branch prediction leads to a remarkable performance improvement (\textit{Opt-6}), reaching 1.30$\times$ for Core i5, and 1.18$\times$ for Xeon Platinum. In the same line, loop unrolling provides good acceleration rates, especially for 
 Core i5 (1.43$\times$). 

\begin{table}[tb]
\centering
\caption{Incremental improvement for each x86 platform when N=8192}
\label{tab:mejoras-incrementales}
\begin{tabular}{@{}ccccccccc@{}}
\toprule
\textbf{x86 platform} & \textbf{Opt-1} & \textbf{Opt-2} & \textbf{Opt-3} & \textbf{Opt-4} & \textbf{Opt-5} & \textbf{Opt-6} & \textbf{Opt-7} & \textbf{Opt-8} \\ \midrule
\textit{Core i5}       & -     & 1.42 & 2.62 & -     & 1.06 & 1.30 & 1.43 & $<$ 1.01 \\
\textit{Xeon Platinum} & -     & 1.32 & 2.61 & 1.15 & 1.21 & 1.18 & 1.09 & 1.13   \\
\textit{Xeon Phi KNL}  & 1.03 & 1.57 & 2.19 & 2.10 & 1.05 & 2.63 & 1.40 & $<$ 1.01 \\ \bottomrule
\end{tabular}
\end{table}

\begin{figure}[tb!]
    \centering
    \includegraphics[width=0.9\columnwidth]{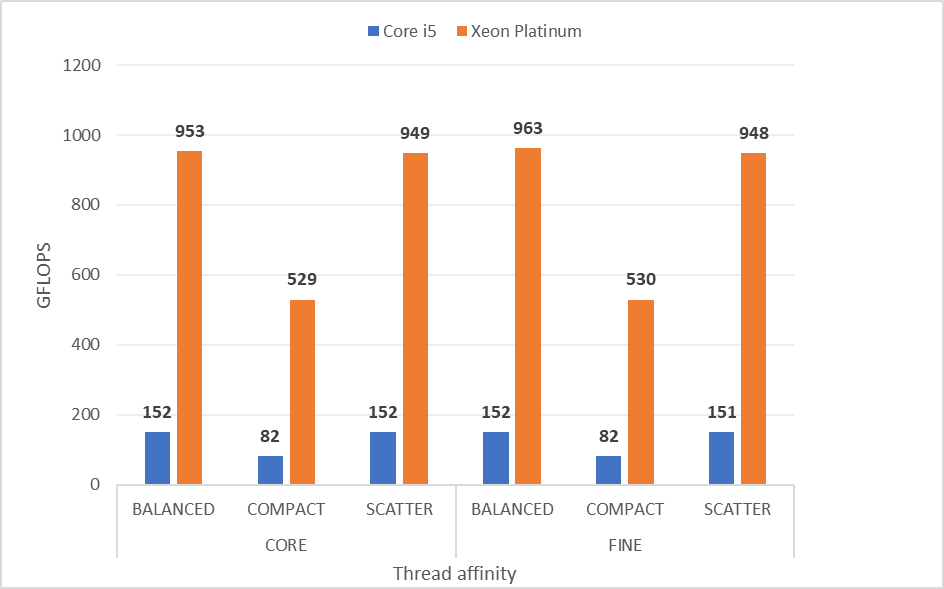}
    \caption{Performance for all combinations of thread affinity on Xeon Platinum and Core i5 when \textit{N}=16384, data type = \textit{float}, and one thread per physical core is set}
    \label{fig:i5-aff}
\end{figure}

For the implementation \textit{Opt-8}, six combinations of granularity (\textit{core, fine}) and affinity (\textit{balanced, compact and scatter}) were tested on both platforms~\cite{AfinidadIntel} (see Fig.~\ref{fig:i5-aff}). The best configuration was \textit{balanced} closely followed by \textit{scatter} on the Xeon Platinum when using one OpenMP thread per physical core. On its behalf, no significant differences are observed between \textit{balanced} and \textit{scatter} on the Core i5, probably due to the small number of available cores. Last, the granularity option does not seem to affect the performance of both machines while the \textit{compact} affinity, on the contrary, affects it negatively.

% TABLA DE MEJORAS INCREMENTALES

\begin{figure}[t]
    \centering
    \includegraphics[width=0.9\columnwidth]{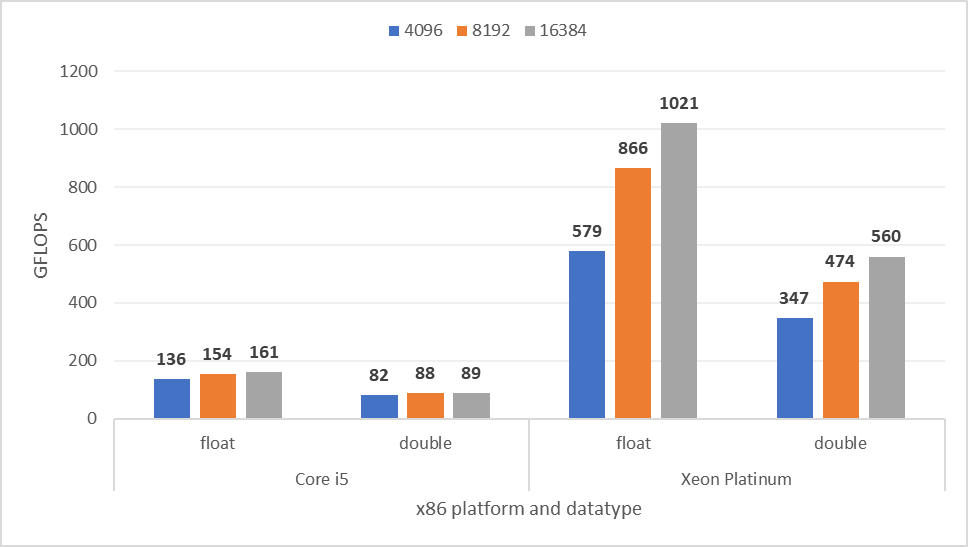}
    \caption{Performance of optimal implementation on Xeon Platinum and Core i5 varying N and data type}
    \label{fig:o8-barras-both}
\end{figure}

%En resumen, si se considera el desempeño integral de las versiones tomando como base a \textit{Opt-0}, al sumar las mejoras obtenidas desde \textit{Opt-2} hasta \textit{Opt-8}, se tiene un acumulado de \textbf{7.31x} y \textbf{6.98x} para Core i5 y Xeon Platinum, respectivamente. Adicionalmente, en la comparación con el Xeon Phi KNL, se puede notar que todas las optimizaciones resultaron beneficiosas, aunque no todas impactaron de la misma manera.

When comparing \textit{Opt-8} to \textit{Opt-0}, an overall improvement of 7.31$\times$ and 6.98$\times$ are reached for Core i5 and Xeon Platinum, respectively. Like the Xeon Phi KNL case, it can be noted that all optimizations were beneficial, although not all of them impacted in the same way.

%\begin{figure}[h]
%    \centering
   % \includegraphics[width=0.85\columnwidth]{figs/fw5-f-56-aff.png}
    %\caption{GFLOPS en Xeon Platinum según KMP\_AFFINITY}
%    \label{fig:xeon-aff}

%\end{figure}

%Por último, en la Fig.~\ref{fig:o8-barras-both} se muestran los GFLOPS alcanzados con la mejor versión para cada equipo, tipo de dato y tamaño de entrada testeado, empleando la configuración óptima de $T$ y $TB$ a cada caso. En primer lugar, resulta claro que se obtienen mayores GFLOPS en el equipo Xeon que en el Core i5, considerando su potencia de cómputo. En segundo lugar, se observa que el rendimiento se incrementa a medida que aumenta $N$, dado a la mayor proporción de cómputo frente a sincronización. En particular, la magnitud de esta diferencia es más notoria cuando se dispone de más hilos (caso Xeon Platinum). En tercer lugar, el uso de un tipo de dato de mayor precisión como \textit{double} puede llevar a un resultado más fiable; sin embargo, se debe tener en cuenta que tendrá un costo en el tiempo de respuesta, ya que el rendimiento decae hasta un 45\%.

Finally, Fig.~\ref{fig:o8-barras-both} shows performance achieved with the best implementation for each platform, data type, and input size tested, using the optimal configuration of $T$ and $BS$. First, higher GFLOPS are obtained in the Xeon server than in the Core i5, considering its computational power. Second, it is observed that performance improves as $N$ increases, given the higher ratio of compute versus synchronization. In particular, the magnitude of this difference is more noticeable when more threads are available (Xeon Platinum case). Third, using a wider precision data type such as \textit{double} can lead to a more reliable result; however, it should be noted that it will come at a cost in response time, as performance drops by as much as 45\%.

% comenté la figura ya que seguramente quedará descartada por repetición y falta de espacio

%\begin{figure}[H]
%    \centering
%    \includegraphics[width=0.8\columnwidth]{figs/fw5-%f-56-aff.png}
%    \caption{GFLOPS en Xeon Platinum según %KMP\_AFFINITY}
%    \label{fig:resultados-aff}
%\end{figure}

\subsection{Experimental results of Opt-9}

%La Tabla~\ref{tab:resultados-completos-o9} muestra los resultados obtenidos para las versiones Opt-9-Sem y Opt-9-Cond en el Xeon Platinum. Se puede observar en ellas que el tamaño de bloque óptimo se encuentra estabilizado en $BS = 128$, independientemente del tipo de dato utilizado, cantidad de hilos o tamaño de entrada. Esto representa una ventaja frente a las versiones anteriores con $BS$ óptimo variable, lo que dificultaba la elección del tamaño adecuado a cada situación.

Table~\ref{tab:resultados-completos-o9} shows the performance for the \textit{Opt-9-Sem} and \textit{Opt-9-Cond} versions on the Xeon Platinum machine. It can be seen that the optimal block size is stabilized at $BS = 128$, regardless of the data type, number of threads, or problem size used. This represents an advantage over previous versions with variable optimal $BS$ since it requires adjustment for each situation.

% \usepackage{booktabs}
% \usepackage{multirow}
\begin{table}[H]
\centering
\caption{Performance results for \textit{Opt-9-Sem} and \textit{Opt-9-Cond} implementations}
\label{tab:resultados-completos-o9}
\begin{tabular}{@{}ccccccc@{}}
\toprule
\multirow{2}{*}{\textbf{N}} &
  \multirow{2}{*}{\textbf{T}} &
  \multirow{2}{*}{\textbf{BS}} &
  \multicolumn{2}{c}{\textbf{GFLOPS (double)}} &
  \multicolumn{2}{c}{\textbf{GFLOPS (float)}} \\  
                       &                      &     & \textbf{Opt-9-Sem} & \textbf{Opt-9-Cond} & \textbf{Opt-9-Sem} & \textbf{Opt-9-Cond} \\ 
\cmidrule(r){1-7}
\multirow{8}{*}{4096}  & \multirow{4}{*}{56}  & 32  & 272.76           & 277.13           & 408.88           & 372.02          \\
                       &                      & 64  & 397.35           & 398.82           & 485.91           & 479.04          \\
                       &                      & 128 & \cellcolor{lightgray}{423.49}           & \cellcolor{lightgray}{423.66}           & \cellcolor{lightgray}{581.04}           & \cellcolor{lightgray}{585.81}          \\ 
                       &                      & 256 & 326.07            & 326.21          & 446.40           & 451.08          \\ \cmidrule(r){4-7}
                       & \multirow{4}{*}{112} & 32  & 245.19           & 252.80           & 357.67           & 350.46          \\
                       &                      & 64  & 352.83           & 353.72           & 484.72           & 476.35          \\
                       &                      & 128 & \cellcolor{lightgray}{402.84}           & \cellcolor{lightgray}{387.35}           & \cellcolor{lightgray}{586.41}           & \cellcolor{lightgray}{594.38}          \\ 
                       &                      & 256 & 220.53           & 201.16           & 315.45           & 319.62          \\ \cmidrule(r){1-7}
\multirow{8}{*}{8192}  & \multirow{4}{*}{56}  & 32  & 317.65           & 333.33           & 469.88           & 497.63          \\
                       &                      & 64  & 494.51           & 499.41           & 639.82           & 738.00          \\
                       &                      & 128 & \cellcolor{lightgray}{550.83}           & \cellcolor{lightgray}{554.76}           & \cellcolor{lightgray}{855.15}           & \cellcolor{lightgray}{857.87}          \\ 
                       &                      & 256 & 424.43           & 427.21           & 657.77           & 657.24          \\ \cmidrule(r){4-7}                       
                       & \multirow{4}{*}{112} & 32  & 315.18           & 326.49           & 436.02           & 483.08          \\
                       &                      & 64  & 466.42           & 468.68           & 623.87           & 688.01          \\
                       &                      & 128 & \cellcolor{lightgray}{585.31}           & \cellcolor{lightgray}{549.29}           & \cellcolor{lightgray}{905.83}           & \cellcolor{lightgray}{909.03}          \\ 
                       &                      & 256 & 316.91           & 297.33           & 480.62           & 499.98          \\ \cmidrule(r){1-7}
\multirow{8}{*}{16384} & \multirow{4}{*}{56}  & 32  & 409.05           & 426.83           & 582.79           & 631.27          \\
                       &                      & 64  & 584.32           & 583.84           & 717.50           & 904.71          \\
                       &                      & 128 & \cellcolor{lightgray}{593.36}           & \cellcolor{lightgray}{590.39}           & \cellcolor{lightgray}{976.98}           & \cellcolor{lightgray}{978.05}          \\
                       &                      & 256 & 456.67           & 455.32           & 750.52           & 754.21          \\ \cmidrule(r){4-7}                       
                       & \multirow{4}{*}{112} & 32  & 414.55           & 429.33           & 534.50           & 627.63          \\
                       &                      & 64  & 556.36           & 541.68           & 706.58           & 828.90          \\
                       &                      & 128 & \cellcolor{lightgray}{629.38}           & \cellcolor{lightgray}{589.54}           & \cellcolor{lightgray}{1034.44}          & \cellcolor{lightgray}{1038.15}         \\ 
                       &                      & 256 & 339.68           & 316.24           & 797.82          & 561.74         \\ 
                       \bottomrule
\end{tabular}
\end{table}

%La Figura~\ref{fig:o9-final} resume los rendimientos obtenidos para ambas versiones en el Xeon Platinum (\textit{T}=112), al variar N y el tipo de datos. En primer lugar, se puede observar que ambas versiones de Opt-9 mejoran el rendimiento de Opt-8. En cuanto al mecanismo de sincronozación, parece no haber diferencias significativas entre usar semáforos o variables condition. Por último, la mejora lograda con tipo de dato double es bastante más grande que con float. En particular, la mejora con float alcanza hasta 5\% mientras que con double se logra hasta un 23\% adicional de GFLOPS. La mejora con double es mayor debido a que al ser más costoso computar con este tipo de dato, proporcionalmente hay menos ocio que con float.

\begin{figure}[t]
    \centering
    \includegraphics[width=0.9\columnwidth]{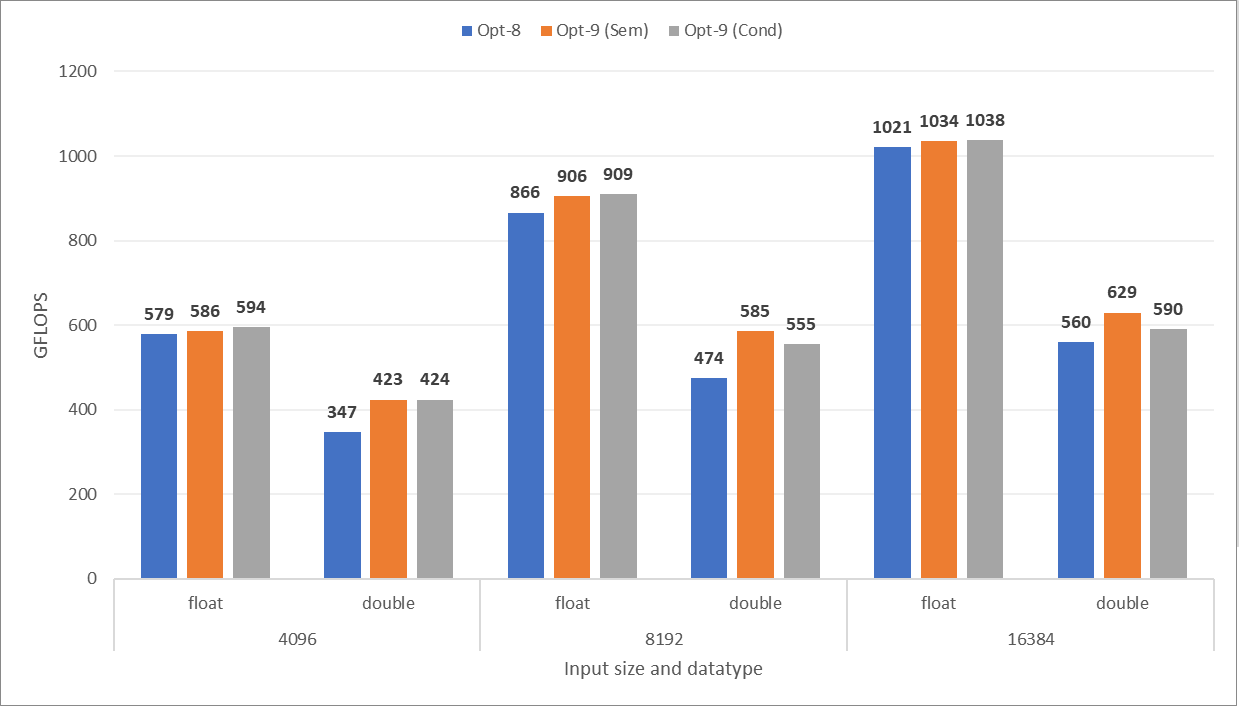}
    \caption{Performance of implementation \textit{Opt-9} on the Xeon Platinum when varying \textit{N} and data type}
    \label{fig:o9-final}
\end{figure}

Fig.~\ref{fig:o9-final} summarizes the performances obtained for both versions on the Xeon Platinum (\textit{T}=112), when varying N and data type. First, it can be seen that both versions of \textit{Opt-9} outperform \textit{Opt-8}. From the synchronization mechanism perspective, no significant performance difference can be appreciated between using semaphores or condition variables. Finally, the improvement achieved with \textit{double} data type is significantly larger than with \textit{float}. 
  In particular, performance improves up to 5\% and 23\% when \textit{float} and \textit{double} are used, respectively. The improvement factor is higher with \textit{double} because idle time is (proportionally) shorter than with \textit{float}, since operations with the former are more expensive than with the latter.

\section{Conclusions and Future Work}
\label{sec:conclusiones}

%En este artículo, se tomó el trabajo realizado por Costi\textit{et al.}~\cite{costi2020aceleracion} y se adaptó su código para que pueda ejecutarse en procesadores multicore Intel x86, abandonando la especificidad del Xeon Phi KNL. Para ello, se verificó una a una las optimizaciones propuestas en el trabajo de base, realizando ajustes a su código donde fuese necesario y analizando su rendimiento en dos servidores de referencia ante diferentes escenarios de prueba. Además, se propuso una nueva optimización para mejorar el grado de  concurrencia del algoritmo paralelo, la cual fue implementada de dos maneras diferentes. A partir de los resultados obtenidos y su posterior análisis, se pueden mencionar las siguientes conclusiones:

In this paper, we studied the code developed by ~\cite{costi2020aceleracion} and adapted it to run on Intel x86 processors, losing the specificity of the Xeon Phi KNL. To do so, we verified one by one the optimizations proposed by~\cite{costi2020aceleracion}, making adjustments to the base code where necessary, and analyzing its performance on two Intel servers under different test scenarios. In addition, a new optimization was proposed to increase the concurrency degree of the parallel algorithm, which was implemented using two different synchronization mechanisms. From the results obtained and their subsequent analysis, the following conclusions can be mentioned:

\begin{itemize}

        \item \textit{Opt-1} was discarded because of the absence of MDCDRAM memory on the x86 platforms used. In addition, the vectorization flags were modified to the corresponding SIMD sets.

\item Like the Xeon Phi KNL case, all optimizations were beneficial on the two x86 platforms selected. Particularly, the use of SIMD instructions provided the greatest performance improvement.

\item The performance improves as $N$ increases, given the higher ratio of compute versus synchronization. Besides, using wider precision data can lead to a more reliable result although at the cost of a significant increase in response time.
        
\item Both versions of \textit{Opt-9} outperform \textit{Opt-8}. From the synchronization mechanism perspective, no significant performance difference can be appreciated between using semaphores or condition variables. In the opposite sense, the improvement achieved with \textit{double} data type is significantly larger than with \textit{float}. 

    \item  Beyond the reduction in execution time, an indirect benefit of \textit{Opt-9} results in no variation of optimal $BS$. Using \textit{Opt-9} the optimal block size is stabilized at $BS = 128$, regardless of the data type, number of threads, or problem size used. This represents an advantage over previous versions with variable optimal $BS$ since it requires adjustment for each situation.
        
\end{itemize}

Future work will focus on:

\begin{itemize}

    \item Proposing new algorithmic optimizations as an inter-round optimization, to remove the synchronization barrier at the end of each round. Then, performing the corresponding tests to evaluate their feasibility.

    \item Making adjustments to the code to enable its compilation using Intel's new ICX compiler, which incorporates LLVM as \textit{backend}. This will guarantee long-term support for the code.

    \item Developing a library to facilitate the inclusion and use of the optimized, parallel FW algorithm in third-party C/C++ programs.
\end{itemize}

%\bigskip \noindent\textbf{Agradecimientos.} Los autores agradecen el soporte de la empresa NVIDIA por la donación de la GPU Titan X usada en esta investigación.

\bibliographystyle{splncs04}
\bibliography{references}
\end{document}